\documentclass[traditabstract]{aa} 
\usepackage{graphicx}
\usepackage{lscape}
\usepackage{natbib}
\usepackage{txfonts}
\begin{document}
   \title{Herschel's view into Mira's head\thanks{{Herschel} is an ESA space observatory with science instruments provided by
European-led Principal Investigator consortia and with important participation from NASA.}}

   \author{A. Mayer
          \inst{1}
          \and A. Jorissen
          \inst{2}
          \and F. Kerschbaum
          \inst{1}
          \and S. Mohamed
          \inst{3}
          \and S. Van Eck
          \inst{2}
          \and R. Ottensamer
          \inst{1}
          \and J.A.D.L. Blommaert
          \inst{4}
          \and L. Decin
          \inst{4}
          \and M.A.T. Groenewegen
          \inst{5}
          \and Th. Posch
          \inst{1}
          \and B. Vandenbussche
          \inst{4}
          \and  Ch. Waelkens
          \inst{4}
          }

\institute{
University of Vienna, Department of Astronomy, T\"urkenschanzstra\ss e 17, A-1180 Wien, Austria\\
 \email{a.mayer@univie.ac.at}
\and
Institut d'Astronomie et d'Astrophysique, Universit\'e Libre de Bruxelles, CP. 226, 
  Boulevard du Triomphe, B-1050 Brussels, Belgium 
\and
Argelander-Institut f\"ur Astronomie, Universit\"at Bonn, Auf dem H\"ugel 71, D-53121 Bonn, Germany
\and
Instituut voor Sterrenkunde, K.U. Leuven, Celestijnenlaan, 200D, B-3001 Leuven, Belgium  
\and
Koninklijke Sterrenwacht van Belgi\"e, Ringlaan 3, B-1180 Brussels, Belgium
}

\date{Received ...; accepted ...}

\abstract
{Herschel's PACS instrument observed the environment of the binary system Mira Ceti in the 70 and 160 $\mu$m bands. These images reveal bright structures shaped as five broken arcs and fainter filaments in the ejected material of Mira's primary star, the famous AGB star $o$ Ceti. The overall shape of the IR emission around Mira deviates significantly from the expected alignment with Mira's exceptionally high space velocity. The observed broken arcs are neither connected to each other nor are they of a circular shape; they stretch over angular ranges of 80 to 100 degrees. By comparing Herschel and GALEX data, we found evidence for the disruption of the IR arcs by the fast outflow visible in both H$\alpha$ and the far UV.  Radial intensity profiles are derived, which place the arcs at distances of 6--85$\arcsec$ (550 -- 8000~AU) from the binary.  Mira's  IR environment appears to be shaped by the complex interaction of Mira's wind with its companion, the bipolar jet, and the ISM. 
}
 
   \keywords{
Stars: oxygen --
Stars: asymptotic giant branch --
Stars: Mass loss --
Stars: Binarity --
Infrared: stars --
Interstellar medium
}

\maketitle
%

\section{Introduction}

Centuries ago $o$ Ceti confused the astronomers all over the world with its appearance and disappearance 
on the night sky and was therefore named \textit{Mira}, the wonderful, by \citet{HEVELIUS1662}.

 Today we know that Mira is a wind-accreting binary star system  \citep{REIMERS1985} at a distance of 91.7~pc \citep[corresponding to a parallax of $10.9\pm1.2$~mas;][]{VANLEEUWEN2007} consisting of the primary star Mira~A ($o$~Ceti) and its companion Mira~B (VZ~Ceti). Mira A is an M-type oxygen-rich star with a pulsation period of 331~d and a brightness amplitude in the visible of 8~mag. Moreover, Mira~A is the prototype of stars in their late stage of evolution residing on the asymptotic giant branch (AGB). The nature of Mira~B has been  unknown for long; with an effective temperature of about 10\ts000~K, it might either be a main-sequence star or a white dwarf \citep{KAROVSKA1997}, but recent observations show strong evidence in favour of a white dwarf \citep{SOKOLOSKI2010}. From a Hubble Space Telescope observation,  \cite{KAROVSKA1997} found  the separation of Mira AB to be $\approx$0.6$\arcsec$, which corresponds to 55~AU.
A very uncertain orbit, based on astrometric measurements since 1923, which has a period of $\approx$500~yr, has been published by \citet{PRIEUR2002}.
\\
Stars on the AGB loose a significant fraction of their mass in the form of a stellar wind. For Mira~A, a moderate mass loss rate of 10$^{-7}$ M$_\odot$ yr$^{-1}$ was determined by \citet{MAURON1992}, which corresponds to a dust mass-loss rate of about $5.5\times10^{-10}$~M$_\odot$~yr$^{-1}$ (Meixner, 2011, priv. comm.). From mm-CO lines, \citet{JOSSELIN2000} found  a wind expansion velocity of 4~km~s$^{-1}$ and a molecular envelope around the binary system with a size of about 20$\arcsec$. This envelope is disrupted by a slow bipolar outflow of about 8~km~s$^{-1}$ \citep{JOSSELIN2000}, but a high-velocity component at 160~km~s$^{-1}$ has been observed by \citet{MEABURN2009}. In the outer environment of Mira, a spectacular observation with the GALEX ultraviolet satellite \citep{MARTIN2007} revealed a cometary-like structure with an extension of more than 2$^\circ$ ($\approx 4$~pc) from head to tail. \citet{MARTIN2007} attribute this tail to Mira's exceptionally high space velocity of $\approx$110~km~s$^{-1}$ and the resulting strong interaction of the stellar wind with the interstellar medium (ISM). This structure was also observed in the 21~cm H~I line \citep{MATTHEWS2008} and is well reproduced by the hydrodynamic simulations of \citet{WAREING2007}, \citet{RAGA2008B} and \citet{ESQUIVEL2010}. In the vicinity of the binary system,  knot-shaped enhancements of the UV and H$\alpha$ emission were found at a distance of 1$\arcmin$ to 2$\arcmin$ from the binary \citep{MARTIN2007,MEABURN2009}, corresponding to 5500 to 11\ts000~AU, and in a direction  consistent with that of the bipolar outflow observed by \citet{JOSSELIN2000}. In the far IR \citet{UETA2008} resolved with the Spitzer Space Telescope the contours of Mira's astropause including the circumbinary envelope, termination shock, and astrosheath.
\\
The present study offers a detailed view  of Mira's circumstellar shell(s)  thanks to the high spatial resolution of the PACS instrument (which has a point-spread function of $5\farcs6$ full-width at half-maximum  at 70~$\mu$m) onboard the Herschel satellite. The images presented here reveal clumpy  broken arcs  (see Sect.~\ref{analysis}) that result from the complex interactions of the wind with the ISM, the companion and the fast outflow.

\begin{figure}[t]
\centering
   \includegraphics[width=4cm]{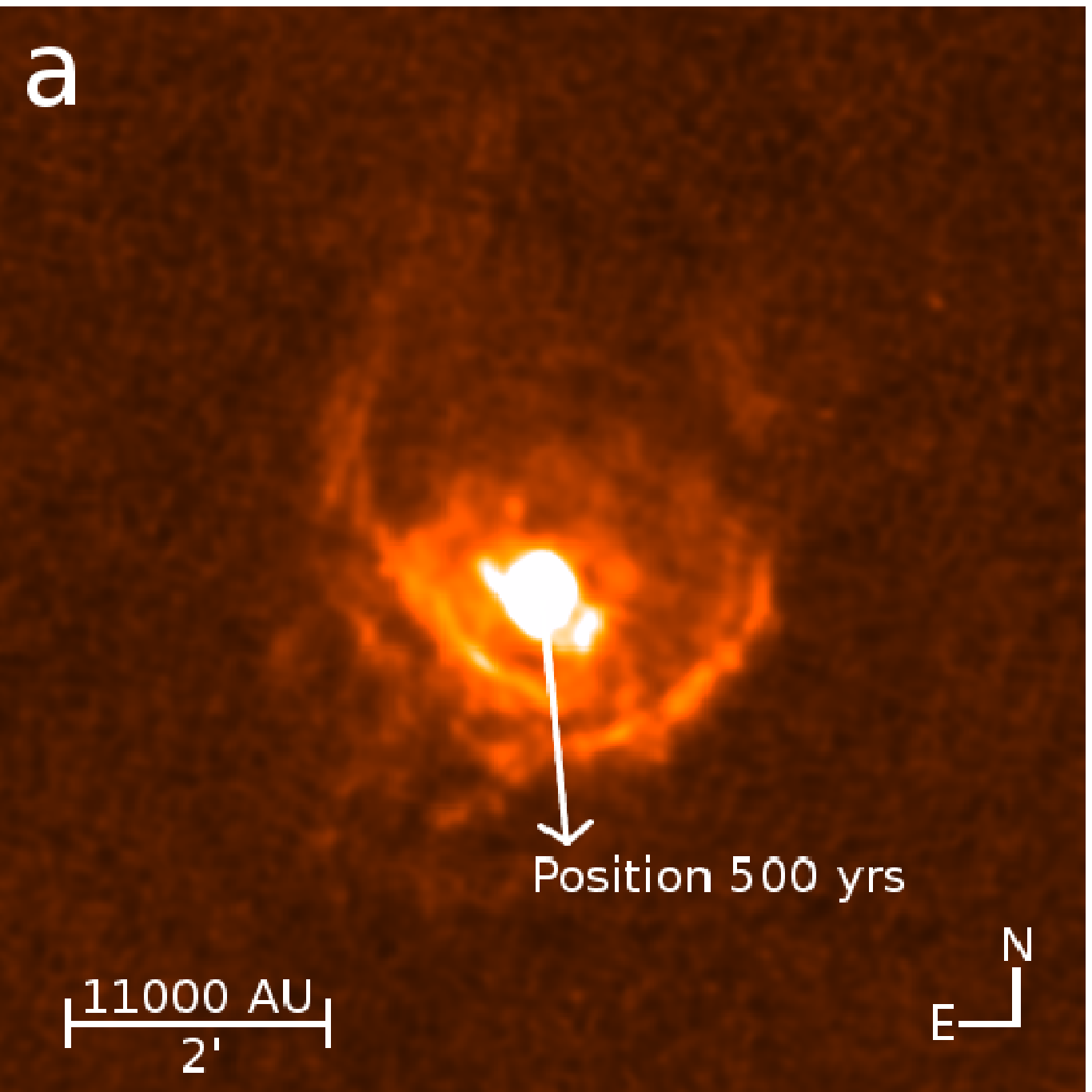}
\includegraphics[width=4.35cm]{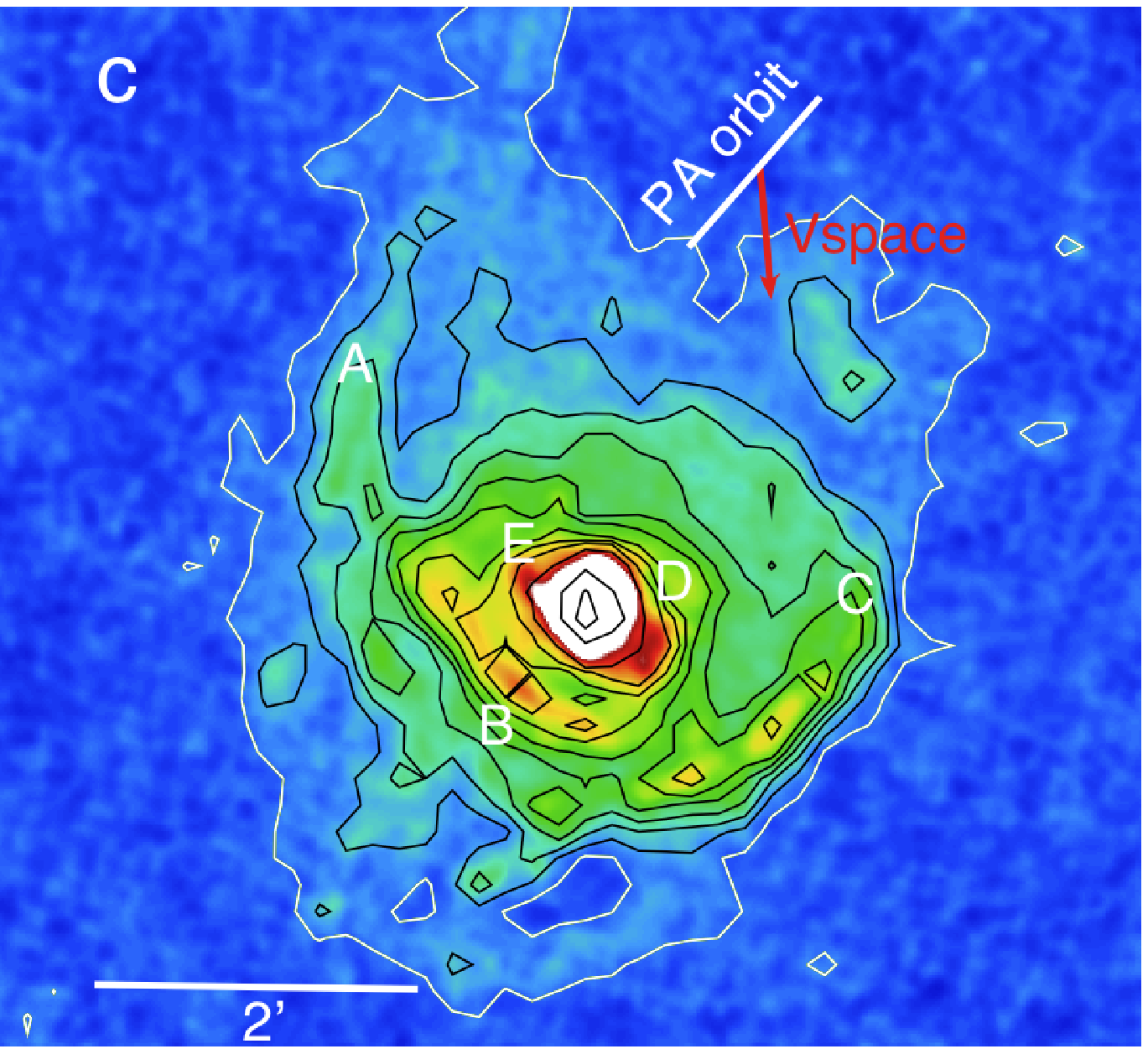}\\
  \includegraphics[width=4cm]{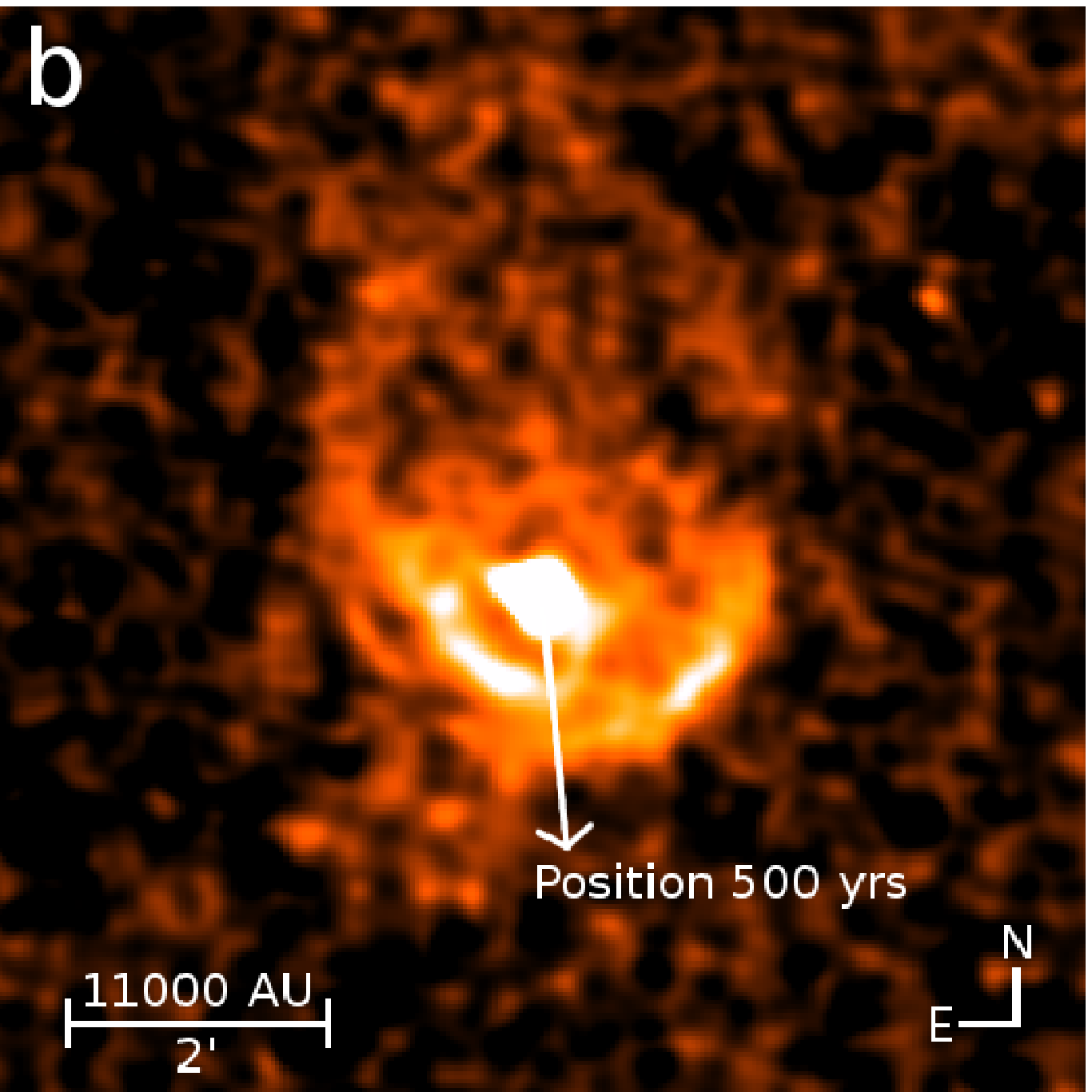}
\includegraphics[width=4.35cm]{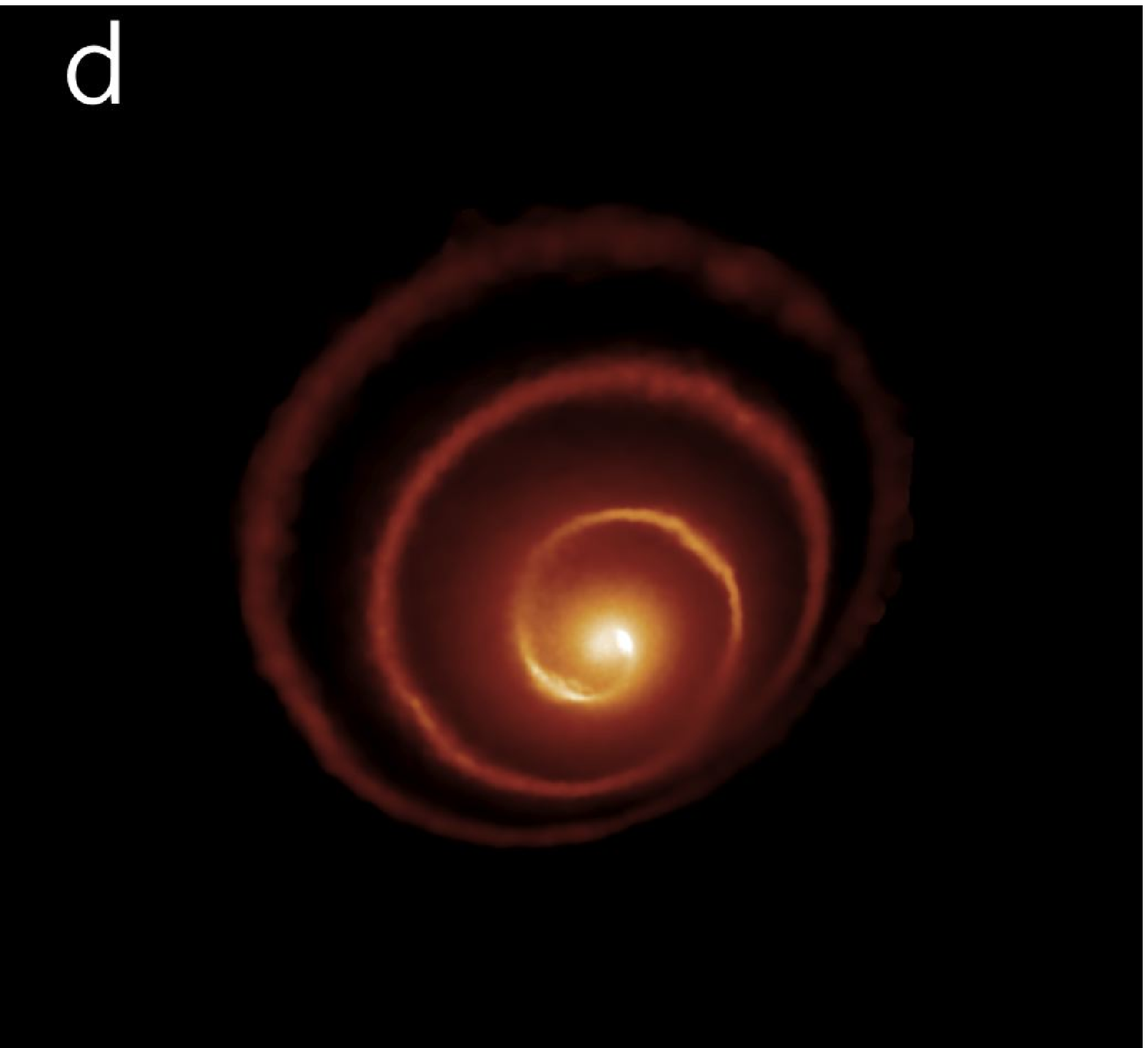}

     \caption{\label{Fig:mira_panel}
Panel (a): deconvolved PACS image at 70~$\mu$m. The arrow indicates the space motion and the position in 500 yrs; (b) is the same for the 160~$\mu$m band and (c) is the 70~$\mu$m deconvolved PACS image with contours and arcs labelled as A, B, C, D and E. The lowest contour is the 3$\sigma$ threshold (and is depicted in white), the arrow labelled with V$_{\rm space}$ shows the direction of the space motion, the bar labelled PA orbit gives the orientation of the major axis of the apparent orbit, orthogonal to the node line.  Panel (d) results from the `toy model' described in Sect.~\ref{model} that is based on the hydrodynamical simulations of \citet{MOHAMED2007,MOHAMED2011}.
}
\end{figure}

\begin{table}
\caption{\label{Tab:data}
Kinematic data for Mira. Negative inclinations are towards the observer. The various vectors are displayed in a 3D view in Fig.~\ref{Fig:3D}. The adopted solar motion to convert heliocentric velocities in LSR velocities is $(U,V,W)_\odot = (11.10, 12.24, 7.25)$~km~s$^{-1}$  \citep{SCHONRICH2010}. The matrix to convert equatorial velocities into galactic velocities is taken from  \citet{JOHNSON1987}. 
}
\begin{tabular}{ll|ll}
\hline
$b (^\circ)$ & $-58$ & 
$\varpi$ (mas) & $10.9\pm1.2^a$ \\
$\mu_{\alpha, LSR}$ (mas yr$^{-1}$)& $-20.93^a$ & 
$\mu_{\delta, LSR}$ (mas yr$^{-1}$) & $-222.39^a$ \\
$RV_{LSR}$ (km s$^{-1}$) & 47.2$^f$ &
$V_{\rm space, LSR}$ (km s$^{-1}$) & 107.9 \\
$V_{\rm space, LSR}$ P.A. $(^\circ)$ & 185.4 &
$V_{\rm space, LSR}$ $i$ $(^\circ)$ & 25.9 \\
$V_{\rm wind}$ (km s$^{-1}$) & 4$^c$ &
$V_{\rm outflow}$ (km s$^{-1}$) & $160\pm10^d$ \\
Bipolar flow PA $(^\circ)$& 168-180$^d$ &  
Bipolar flow $i$ $(^\circ)$ &  -69$^d$ \\
Orbital pole PA $(^\circ)$ & 228$^e$ & 
Orbital pole $i$ $(^\circ)$ & $\pm22^e$ \\
$P$ (yr)   & $\approx$500$^e$ &
$a$ (AU) & $\approx$100$^{a,e}$ \\
$M_1+M_2$ (M$_\odot$) & 4.4$^e$ & 
$\dot{M}$ (M$_\odot$ yr$^{-1}$) & 10$^{-7}$ $^b$ \\
\hline
\end{tabular}

a: \citet{VANLEEUWEN2007}; b: \citet{MAURON1992}; c:  \citet{JOSSELIN2000}; d: \citet{MARTIN2007,MEABURN2009}; e: \citet{PRIEUR2002}; f: \citet{FONG2006}
\end{table}

\section{Observations and data reduction}
\label{observations}

The observations of Mira took place on February 9, 2010, using the PhotoArray Camera and Spectrometer \citep[PACS;][]{POGLITSCH2010} onboard Herschel as part of the MESS (Mass-loss of Evolved StarS) Guaranteed Time Key Program \citep{GROENEWEGEN2011}.

The PACS camera observed at 70~$\mu$m and 160~$\mu$m simultaneously, 
covering a field of $30\arcmin \times 30\arcmin$ on the sky with a scan-map mode. 
For the basic data processing  to produce the images displayed in Fig.~\ref{Fig:mira_panel}, the Herschel Interactive Processing Environment (HIPE) was used \citep{ARDILA2010} together with Scanamorphos, an IDL software for building scan maps from PACS (and SPIRE) observations \citep{ROUSSEL2010}. Finally, we deconvolved the images as described by \citet{OTTENSAMER2011}. Although the PACS instrument offers a resolution of 3.2$\arcsec$ per pixel in the 70~$\mu$m (blue) band and 6.4$\arcsec$ per pixel at 160~$\mu$m (red band), the images shown in Fig.~\ref{Fig:mira_panel} are oversampled by a factor 3.2, resulting in a sampling  of 1$\arcsec$ and 2$\arcsec$ per pixel (at 70 and 160~$\mu$m, respectively), to be compared with the  FWHM of $5\farcs6$  and $11\farcs4$ at those wavelengths. We estimate the average background emission to be $1.5\times10^{-4}$~Jy~pixel$^{-1}$ in the blue and $3\times10^{-4}$~Jy~pixel$^{-1}$ in the red band, yielding the 1$\sigma$ sensitivity. 


\section{Results and discussion}

\subsection{IR images}
\label{analysis}

\begin{figure}
\centering
   \includegraphics[width=9cm]{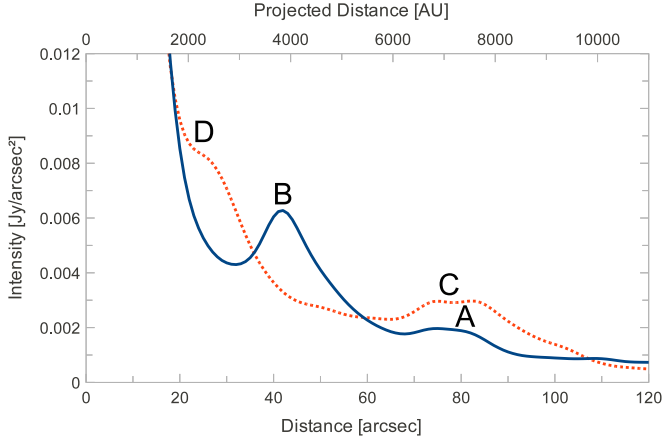}
     \caption{\label{Fig:intensity_plots}
Radial intensity profiles of Mira in the 70~$\mu$m band as a function of the distance to the star. Continuous (blue) line: integrated intensity across the arcs B and A ($75^{\circ} \le PA \le 180^{\circ}$). Dashed (red) line: across the arcs D and C ($190^{\circ} \le PA \le 270^{\circ}$). 
The top scale gives the distance from the star computed from Mira's distance of 91.7~pc.}
\end{figure}

Mira's environment is remarkably structured and exhibits arc-like structures in the surrounding of the binary system, labelled A to E in Fig.~\ref{Fig:mira_panel}c. The best-defined and brightest arc, C, stretches over position angles (PAs) $\sim 190^\circ - 270^\circ$ (south-west to west), is well detached from the star and covers a range of angular separation from $67\arcsec$ (6100~AU) at the closest point to $85\arcsec$ (7800~AU) at the most distant point from the star. The intensity was integrated over this PA range  and the resulting radial intensity profile (Fig.~\ref{Fig:intensity_plots}) clearly shows the arc  as a plateau in the diagram. The extension of this plateau  confirms that the arc is non-circular or, if circular, is not centred on the position of the star. 
We found another arc, D, in the same direction as arc~C but much closer to the star and stretching over much smaller PAs. In the radial intensity profile, arc~D is visible as a bump located at a distance of 21--$28\arcsec$ (1900--2600~AU).
East of Mira more arcs are visible. In the south-east direction, we see arc~B close to the star. It stretches over PAs 75$^\circ$ to 180$^\circ$ and is more circular than arc~C  and therefore appears as a narrow peak in the  radial intensity profile at a distance of 41$\arcsec$ (3800~AU) from the binary (Fig.~\ref{Fig:intensity_plots}).
\\
Farther out, directly to the east, we find another structure, arc~A, stretching towards north-east. In the integrated intensity profile, it can be recognized as a small plateau representing its fairly faint nature at a distance of 68--$89\arcsec$ (6200--8200~AU). This matches the distance of arc~C, which leads to the suggestion that both arcs are somehow related to each other.
\\
It is also possible that another arc, E, is present in the south-east direction, very close to the binary at a distance of less than $15\arcsec$ (1400 AU), of which we only see the edge at a PA of about 65$^\circ$.
It is noteworthy that the arc sequence E-D-B-C-A is a sequence of both increasing distance from the star and increasing PA range, arc E being the shortest and closest to the star, and A the longest and farthest.
\\
The overall 3$\sigma$ emission above the 70~$\mu$m IR background of Mira, visible in Fig.\ref{Fig:mira_panel}c as the outermost contour, spans as far as $157\arcsec$ towards the south-east. It is remarkable that  the extension of the emission is towards a position angle of $\approx 160^\circ$ because one would expect the emission to be aligned with Mira's space motion, but that motion is in the direction of PA = 185$^{\circ}$. In Fig.~\ref{Fig:mira_panel}c we also indicated the orientation of the major axis of the apparent orbit, which is oriented along a PA of $\approx$140$^\circ$ \citep[][]{PRIEUR2002}. As discussed in more detail in Sect.~\ref{model}, the orbital motion and the mass transfer from the binary can also shape its environment. Nevertheless, the extension of the IR emission lies between both values and makes a distinct attribution difficult.

\subsection{Comparison with GALEX UV observations}
\label{galex}

\begin{figure}
\centering
  \includegraphics[width=9cm]{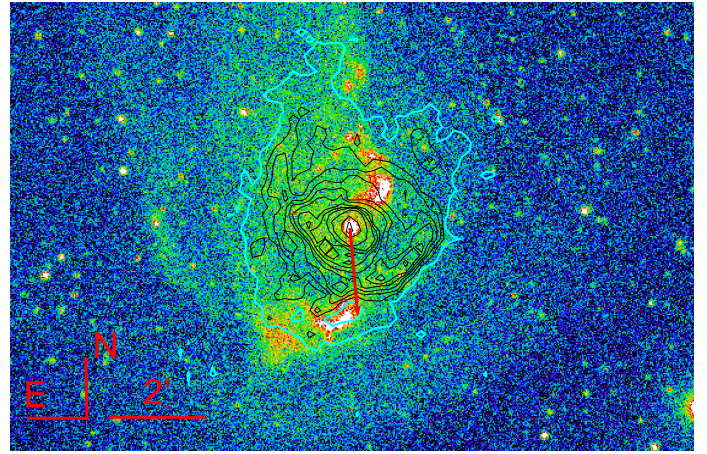}
     \caption{\label{Fig:Galex}
     GALEX far UV image of Mira obtained by \citet{MARTIN2007} overplotted with contours of the Herschel 70 $\mu$m image, the thick blue contour
corresponding to an IR flux just 3$\sigma$ above background. The bright white regions 
correspond to strong UV emission at both ends of the fast jet. Note the absence of IR emission at the end of the southern jet.
The red arrow indicates the position of the star in 500 yrs. The bow shock is the faint structure in the background in the south-west. 
} 
\end{figure}

Before drawing conclusions from the structures observed in the IR, it is interesting to confront these IR observations with GALEX ultraviolet observations of Mira 
in the far and near UV \citep[152.8~nm and 227.1~nm, respectively;][]{MARTIN2007}.
In the far UV image, \citet{MARTIN2007} found emission knots arranged north-west  and south (Fig.~\ref{Fig:Galex}). The axis of the streams is consistent with the direction of the bipolar outflow detected by \citet{JOSSELIN2000}, and moreover, the UV knots are exactly coincident with similar knots seen in H$_\alpha$ \citep{MEABURN2009}.
In contrast, Fig.~\ref{Fig:Galex} reveals that in the PACS 70~$\mu$m image the southern  stream exactly corresponds to a region with an IR flux below the 3$\sigma$ threshold\footnote{The `hole' in the IR emission located about 1$\arcmin$ to the East is less significant than the one considered, with fluxes about twice smaller for the latter.}. 
It thus looks as if the southern stream is penetrating and disrupting the material responsible for the IR emission. According to \citet{MEABURN2009}, the southern stream is indeed inclined at an angle of 69$^\circ$ with respect to the plane of sky  towards us. Put together, all these arguments clearly indicate that the IR emission must come {\it from a 3D structure} and is not confined to the orbital plane; otherwise, there could be no explanation for the anti-coincidence between the GALEX far UV/H$\alpha$ emission and the IR emission. 
\\
Another interesting clue is that the northern lobe of the bipolar flow remains visible even though it is located behind the IR shell. The dust present in the shell will efficiently absorb the far UV radiation, so that the intensity of the northern lobe measured by GALEX should be smaller than that of the southern lobe, which has no intervening dust. 
In the far UV, the peak intensity of the northern  lobe is indeed about 1.5 times fainter than the peak intensity of the southern lobe. Assuming that the northern and southern jets carry the same energy, this ratio may then be seen as resulting from dust extinction:
\begin{equation}
\label{Eq:ext}
1/1.5 = {\rm exp}\left( - N \pi a^2 Q_{\rm ext, UV} \right),
\end{equation}
where $N$ is the dust column density, $a$ is the grain radius, and $Q_{\rm ext,UV}$ is the ratio between the extinction cross section and the geometrical section. This constraint can be inserted in Kirchhoff's law for the IR emission $j_{IR}$,  leading to
\begin{equation}
j_{\rm IR} = B_{\rm IR}(T_{\rm dust})\;\ln(1.5)\;  Q_{\rm abs,IR} / Q_{\rm ext,UV} ,
\end{equation}
where $B_{\rm IR}$ is the Planck intensity function at the dust temperature and $Q_{\rm abs,IR}$ the ratio of the absorption cross section and the geometrical section. From \citet{DRAINE1985}, $Q_{\rm abs, 70\mu m} / Q_{\rm ext, 0.15 \mu m} \sim 10^{-3}$ (almost independent of the grain size), which leads to
\begin{equation}
\label{Eq:prediction}
j_{\rm IR} = 4.2 \times 10^{-4}\;  B_{\rm IR}(T_{\rm dust}).
\end{equation}
From the measured intensity ratio at 70 and 160~$\mu$m for the northern lobe, we find a dust temperature of 36~K; combining $B_{\rm 70\mu m}(36~{\rm K})$ with the measured  $j_{\rm 70\mu m} = 4.42\times10^{7}$~Jy/sr, we obtain $j_{\rm 70\mu m} / B_{\rm 70\mu m}(36~{\rm K}) = 1.2 \times 10^{-4}$, iwhich  agrees well with the value predicted by  Eq.~\ref{Eq:prediction}.

\subsection{Role of the binary system}
\label{model}

The arcs observed around Mira (Sect.~\ref{analysis}) are (almost) unique in the MESS sample\footnote{The prototypical carbon star 
IRC~+10216 is surrounded by a large number of thin and faint arcs, known for quite some time  in the optical  \citep{MAURON1999,MAURON2000}, and now identified in Herschel images as well (Decin et al., in preparation). Their appearance is, however, quite different from the few thick arcs seen around Mira, in line with the fact that there is so far no evidence that IRC~+10216 belongs to a binary system \citep{HUGGINS2009}.}   \citep[][Cox et al., in preparation]{GROENEWEGEN2011}, which makes it tempting to relate them to the binary nature of Mira. 

\begin{figure}[h]
\vspace*{-1cm}
\hspace*{-1.2cm}
\includegraphics[width=7cm,angle=-90]{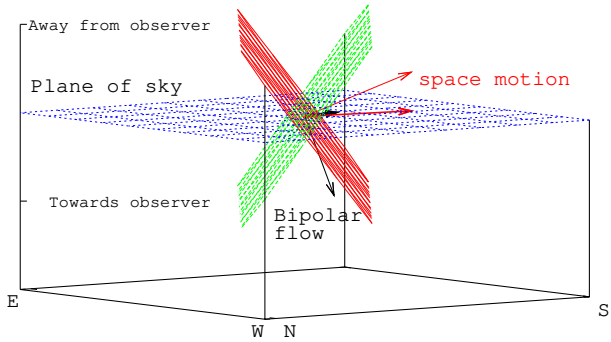}
\vspace*{-1cm}
\caption{Various vectors entering the discussion in a 3D view, with the red and green planes corresponding to the two possible orbital orientations. Because bipolar flows are generally perpendicular to the orbital plane, the green orientation seems the most likely. The  sky plane is depicted in blue. The thick arrows are projections onto the sky plane. The observer is located on the bottom plane and looking upwards.}
\label{Fig:3D}
 \end{figure}

In binary systems involving a mass-losing component, nested half-shells are seen  when the orbit is orientated close to edge-on \citep{HE2007}, and they represent the 3D structure extending above the Archimedian spiral present in the plane of a circular orbit (for eccentric orbits, the spiral becomes broken).
The latter was predicted by the hydrodynamics simulations of  \citet{THEUNS1993},   
\citet{MASTRODEMOS1999}, \citet{MOHAMED2011}, and detected around AFGL~3068 by \citet{MAURON2006} and \citet{MORRIS2006}.
In Mira, the observed arc structure resembles neither an Archimedian spiral, nor nested half-shells; its interpretation must  therefore be more complex, involving  two supplementary physical processes at play, namely the presence of a bipolar jet, and the interaction with the ISM (to fix the ideas, Fig.~\ref{Fig:3D} presents the respective orientations of all relevant vectors involved in the Mira system). We already argued in Sect.~\ref{galex} that the fast bipolar outflow  \citep[160~km~s$^{-1}$;][]{MARTIN2007,MEABURN2009} seems to cut a hole in the IR shell  (see Fig.~\ref{Fig:Galex}). Similarly, it could easily disrupt the arcs with its high velocity, and may actually be responsible for cutting the edges of arcs B, C, and D, because all these edges are located along the southern jet,  at a position angle close to 180$^\circ$.

On the other hand, it is likely that the interaction with the ISM, more precisely, the termination shock, plays some role in shaping (at least) arc C, as already suggested by \citet{UETA2008} from lower-resolution {\it Spitzer} images, and confirmed in the {\it Herschel} images by the sharp ridge  seen on the contour levels in Figs.~\ref{Fig:mira_panel}c  and \ref{Fig:Galex}. Arc~C is indeed located almost upstream along Mira's space motion. On the 160~$\mu$m image (Fig.~\ref{Fig:mira_panel}b),  arcs A and C extend towards the north and join with each other, exactly opposite to the space motion of Mira, in a tail-like or bullet-like structure, as seen in the hydrodynamic simulations by \citet{WAREING2007} and \citet{ESQUIVEL2010}. This is clear evidence that the outer arcs feel the space motion and interact with the ISM. Another indication thereof is provided  by the curvature of the bipolar-outflow southern stream, which  \citet{MARTIN2007} attribute to its deceleration  by the post-shock flow. Therefore, the arcs seen around Mira likely result from a combination of the projected 3D structures resulting from the interaction of Mira's wind with its companion on one hand, and with the ISM on the other hand.  

As a first exploration of this combination, we used the simulations of  \citet{MOHAMED2007,MOHAMED2011} of an AGB star nearly filling its Roche lobe in a system 10~AU-wide\footnote{Although this separation is much smaller than the actual orbital separation of Mira system, this difference should not impact our conclusions, because the spiral properties are largely scale-invariant.}, and losing mass through a slow wind. The 3D structure resulting from the expanding wind disturbed by the companion's gravitational pull was then intersected with a parabolo\"\i d representing the bow shock. All particles within a thin shell along the parabolo\"\i d  surface were kept and the resulting particle density projected onto the sky plane (the respective orientations of the parabolo\"\i d, the orbital plane, the sky plane, and the companion were taken into account, as represented in Fig.~\ref{Fig:3D}). If the IR emission observed by Herschel around Mira results from the reheating of the wind by the shock, this `toy model', presented in Fig.~\ref{Fig:mira_panel}d, should then give  a fair account of the situation. Fig.~\ref{Fig:mira_panel}d shows that there are indeed promising resemblances between the Herschel image and the one from the `toy model', apart from the absence of brightening of the south-west arc C (where the compression from the ISM -- not included in our model -- is maximum), and apart from the ISM-swept appearance of  the outermost northern arc. Hydrodynamical simulations of mass transfer in the Mira system, {\it including} the ISM interaction,  are thus badly needed.

\begin{acknowledgements}
This work was supported in part by the Belgian Federal
Science Policy Office via the PRODEX Programme of ESA (Nos. C90371 and C90372).  AM and FK acknowledge
funding by the Austrian Science Fund FWF under project number P23586-N16, RO under project number I163-N16.  We thank A. Zijlstra and D. Pourbaix 
for discussions about Mira's orbit. SvE is FNRS Research Associate.
\end{acknowledgements}
\bibliographystyle{aa}
\bibliography{17203}

\begin{thebibliography}{34}
\expandafter\ifx\csname natexlab\endcsname\relax\def\natexlab#1{#1}\fi

\bibitem[{{Ardila} \& {Science Ground Segment Consortium}(2010)}]{ARDILA2010}
{Ardila}, D.~R. \& {Science Ground Segment Consortium}, H. 2010, BAAS, 42, 397

\bibitem[{{Draine}(1985)}]{DRAINE1985}
{Draine}, B.~T. 1985, \apjs, 57, 587

\bibitem[{{Esquivel} {et~al.}(2010){Esquivel}, {Raga}, {Cant{\'o}},
  {Rodr{\'{\i}}guez-Gonz{\'a}lez}, {L{\'o}pez-C{\'a}mara}, {Vel{\'a}zquez}, \&
  {De Colle}}]{ESQUIVEL2010}
{Esquivel}, A., {Raga}, A.~C., {Cant{\'o}}, J., {et~al.} 2010, \apj, 725, 1466

\bibitem[{{Fong} {et~al.}(2006){Fong}, {Meixner}, {Sutton}, {Zalucha}, \&
  {Welch}}]{FONG2006}
{Fong}, D., {Meixner}, M., {Sutton}, E.~C., {Zalucha}, A., \& {Welch}, W.~J.
  2006, \apj, 652, 1626

\bibitem[{{Groenewegen} {et~al.}(2011){Groenewegen}, {Waelkens}, {Barlow},
  {Kerschbaum}, {Garcia-Lario}, {Cernicharo}, {Blommaert}, {Bouwman}, {Cohen},
  {Decin}, {Exter}, {Gear}, {Gomez}, {Hargrave}, {Henning}, {Hutsem\'ekers},
  {Ivison}, {Jorissen}, {Krause}, {Ladjal}, {Leeks}, {Lim}, {Naz\'e},
  {Olofsson}, {Polehampton}, {Posch}, {Rauw}, {Royer}, {Sibthorpe}, {Swinyard},
  {Ueta}, {Vamvatira-Nakou}, {Vandenbussche}, {Van de Steene}, {Van Eck}, {van
  Hoof}, {Van Winckel}, {Verdugo}, \& {Wesson}}]{GROENEWEGEN2011}
{Groenewegen}, M.~A.~T., {Waelkens}, C., {Barlow}, M., {et~al.} 2011, \aap,
  526, A162

\bibitem[{{He}(2007)}]{HE2007}
{He}, J.~H. 2007, \aap, 467, 1081

\bibitem[{{Hevelius}(1662)}]{HEVELIUS1662}
{Hevelius}, J. 1662, {Mercurius in Sole visus Gedani [...] Quibus accedit
  succinta Historiola, nov\ae{} illius, ac mir\ae{} stell\ae{} in collo ceti,
  certis anni temporibus clare admodum affulgentis, rursus omnino evanescentis.
  [...] (Simon Reiniger, Danzig), p.146 sqq.}

\bibitem[{{Huggins} {et~al.}(2009){Huggins}, {Mauron}, \&
  {Wirth}}]{HUGGINS2009}
{Huggins}, P.~J., {Mauron}, N., \& {Wirth}, E.~A. 2009, \mnras, 396, 1805

\bibitem[{{Johnson} \& {Soderblom}(1987)}]{JOHNSON1987}
{Johnson}, D.~R.~H. \& {Soderblom}, D.~R. 1987, \aj, 93, 864

\bibitem[{{Josselin} {et~al.}(2000){Josselin}, {Mauron}, {Planesas}, \&
  {Bachiller}}]{JOSSELIN2000}
{Josselin}, E., {Mauron}, N., {Planesas}, P., \& {Bachiller}, R. 2000, \aap,
  362, 255

\bibitem[{{Karovska} {et~al.}(1997){Karovska}, {Hack}, {Raymond}, \&
  {Guinan}}]{KAROVSKA1997}
{Karovska}, M., {Hack}, W., {Raymond}, J., \& {Guinan}, E. 1997, \apjl, 482,
  L175

\bibitem[{{Martin} {et~al.}(2007){Martin}, {Seibert}, {Neill}, {Schiminovich},
  {Forster}, {Rich}, {Welsh}, {Madore}, {Wheatley}, {Morrissey}, \&
  {Barlow}}]{MARTIN2007}
{Martin}, D.~C., {Seibert}, M., {Neill}, J.~D., {et~al.} 2007, \nat, 448, 780

\bibitem[{{Mastrodemos} \& {Morris}(1999)}]{MASTRODEMOS1999}
{Mastrodemos}, N. \& {Morris}, M. 1999, \apj, 523, 357

\bibitem[{{Matthews} {et~al.}(2008){Matthews}, {Libert}, {G{\'e}rard}, {Le
  Bertre}, \& {Reid}}]{MATTHEWS2008}
{Matthews}, L.~D., {Libert}, Y., {G{\'e}rard}, E., {Le Bertre}, T., \& {Reid},
  M.~J. 2008, \apj, 684, 603

\bibitem[{{Mauron} \& {Caux}(1992)}]{MAURON1992}
{Mauron}, N. \& {Caux}, E. 1992, \aap, 265, 711

\bibitem[{{Mauron} \& {Huggins}(1999)}]{MAURON1999}
{Mauron}, N. \& {Huggins}, P.~J. 1999, \aap, 349, 203

\bibitem[{{Mauron} \& {Huggins}(2000)}]{MAURON2000}
{Mauron}, N. \& {Huggins}, P.~J. 2000, \aap, 359, 707

\bibitem[{{Mauron} \& {Huggins}(2006)}]{MAURON2006}
{Mauron}, N. \& {Huggins}, P.~J. 2006, \aap, 452, 257

\bibitem[{{Meaburn} {et~al.}(2009){Meaburn}, {L{\'o}pez}, {Boumis}, {Lloyd}, \&
  {Redman}}]{MEABURN2009}
{Meaburn}, J., {L{\'o}pez}, J.~A., {Boumis}, P., {Lloyd}, M., \& {Redman},
  M.~P. 2009, \aap, 500, 827

\bibitem[{{Mohamed} \& {Podsiadlowski}(2007)}]{MOHAMED2007}
{Mohamed}, S. \& {Podsiadlowski}, P. 2007, in ASP Conf. Ser. Vol. 372, ed.
  {R.~Napiwotzki \& M.~R.~Burleigh}, 15th European Workshop on White Dwarfs,
  397

\bibitem[{{Mohamed} \& {Podsiadlowski}(2011)}]{MOHAMED2011}
{Mohamed}, S. \& {Podsiadlowski}, P. 2011, in Asymmetric Planetary Nebulae V,
  ed. A.~{Zijlstra}, F.~{Lykou}, I.~{McDonald}, \& E.~{Lagadec} (Jodrell Bank
  Centre for Astrophysics)

\bibitem[{{Morris} {et~al.}(2006){Morris}, {Sahai}, {Matthews}, {Cheng}, {Lu},
  {Claussen}, \& {S{\'a}nchez-Contreras}}]{MORRIS2006}
{Morris}, M., {Sahai}, R., {Matthews}, K., {et~al.} 2006, in IAU Symp., Vol.
  234, Planetary Nebulae in our Galaxy and Beyond, ed. {M.~J.~Barlow \&
  R.~H.~M{\'e}ndez}, 469

\bibitem[{{Ottensamer} {et~al.}(2011){Ottensamer}, {Luntzer}, {Mecina},
  {Kerschbaum}, {Lehmann}, {Blommaert}, {Decin}, {Groenewegen}, {Posch},
  {Vandenbussche}, \& {Waelkens}}]{OTTENSAMER2011}
{Ottensamer}, R., {Luntzer}, A., {Mecina}, M., {et~al.} 2011, in Why galaxies
  care about AGB stars (II) ?, ASP Conf. Ser., in press

\bibitem[{{Poglitsch} {et~al.}(2010){Poglitsch}, {Waelkens}, {Geis},
  {Feuchtgruber}, {Vandenbussche}, {Rodriguez}, {Krause}, {Renotte}, {van
  Hoof}, {Saraceno}, {Cepa}, {Kerschbaum}, {Agn{\`e}se}, {Ali}, {Altieri},
  {Andreani}, {Augueres}, {Balog}, {Barl}, {Bauer}, {Belbachir}, {Benedettini},
  {Billot}, {Boulade}, {Bischof}, {Blommaert}, {Callut}, {Cara}, {Cerulli},
  {Cesarsky}, {Contursi}, {Creten}, {De Meester}, {Doublier}, {Doumayrou},
  {Duband}, {Exter}, {Genzel}, {Gillis}, {Gr{\"o}zinger}, {Henning},
  {Herreros}, {Huygen}, {Inguscio}, {Jakob}, {Jamar}, {Jean}, {de Jong},
  {Katterloher}, {Kiss}, {Klaas}, {Lemke}, {Lutz}, {Madden}, {Marquet},
  {Martignac}, {Mazy}, {Merken}, {Montfort}, {Morbidelli}, {M{\"u}ller},
  {Nielbock}, {Okumura}, {Orfei}, {Ottensamer}, {Pezzuto}, {Popesso},
  {Putzeys}, {Regibo}, {Reveret}, {Royer}, {Sauvage}, {Schreiber}, {Stegmaier},
  {Schmitt}, {Schubert}, {Sturm}, {Thiel}, {Tofani}, {Vavrek}, {Wetzstein},
  {Wieprecht}, \& {Wiezorrek}}]{POGLITSCH2010}
{Poglitsch}, A., {Waelkens}, C., {Geis}, N., {et~al.} 2010, \aap, 518, L2

\bibitem[{{Prieur} {et~al.}(2002){Prieur}, {Aristidi}, {Lopez}, {Scardia},
  {Mignard}, \& {Carbillet}}]{PRIEUR2002}
{Prieur}, J.~L., {Aristidi}, E., {Lopez}, B., {et~al.} 2002, \apjs, 139, 249

\bibitem[{{Raga} \& {Cant{\'o}}(2008)}]{RAGA2008B}
{Raga}, A.~C. \& {Cant{\'o}}, J. 2008, \apjl, 685, L141

\bibitem[{{Reimers} \& {Cassatella}(1985)}]{REIMERS1985}
{Reimers}, D. \& {Cassatella}, A. 1985, \apj, 297, 275

\bibitem[{{Roussel}(2011)}]{ROUSSEL2010}
{Roussel}, H. 2011, \aap, submitted

\bibitem[{{Sch{\"o}nrich} {et~al.}(2010){Sch{\"o}nrich}, {Binney}, \&
  {Dehnen}}]{SCHONRICH2010}
{Sch{\"o}nrich}, R., {Binney}, J., \& {Dehnen}, W. 2010, \mnras, 403, 1829

\bibitem[{{Sokoloski} \& {Bildsten}(2010)}]{SOKOLOSKI2010}
{Sokoloski}, J.~L. \& {Bildsten}, L. 2010, \apj, 723, 1188

\bibitem[{{Theuns} \& {Jorissen}(1993)}]{THEUNS1993}
{Theuns}, T. \& {Jorissen}, A. 1993, \mnras, 265, 946

\bibitem[{{Ueta}(2008)}]{UETA2008}
{Ueta}, T. 2008, \apjl, 687, L33

\bibitem[{{van Leeuwen}(2007)}]{VANLEEUWEN2007}
{van Leeuwen}, F. 2007, \aap, 474, 653

\bibitem[{{Wareing} {et~al.}(2007){Wareing}, {Zijlstra}, {O'Brien}, \&
  {Seibert}}]{WAREING2007}
{Wareing}, C.~J., {Zijlstra}, A.~A., {O'Brien}, T.~J., \& {Seibert}, M. 2007,
  \apjl, 670, L125

\end{thebibliography}

\end{document}